# A Ride to Exoplanets*

## Finding Potentially Habitable Worlds

*Simran Kaur and Varinderjit Kaur*

**Cosmos has always sparked human curiosity to unveil and speculate its fascinating secrets. This curiosity has ultimately opened a window to other worlds. After years of observation, computation, and data analysis, scientists have revealed the diversity in exoplanets that have been helpful in the characterization and further investigation of biosignatures and technosignatures. This article presents some of the scientific advances made in extraterrestrial planetary science that guarantees to search through the thick clouds of the mysterious planets in the near future.**

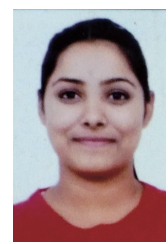

**Simran Kaur** is a final year undergraduate student pursuing major in physics at G.S.S.D.G.S. Khalsa College Patiala. Her interests are Astronomy and Astrophysics.

## 1. Introduction

Who are we? What are we? Why are we here? Are we alone? Are we one of a kind? Or are we just a mere member of a rather larger and yet undiscovered family? These questions have existed for ages. However, there are no concrete answers. But the curiosity to find answers to these problems compelled us to look beyond our own little family—the solar system. The ride to hunt planets marked its starting point in the 1580s [1], accelerating in the 1980s, and one fine day of January 1992, the existence of two exoplanets, the first-ever planets outside our solar system, was confirmed [2].

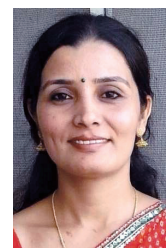

**Varinderjit Kaur** works at G.S.S.D.G.S. Khalsa College Patiala. She is interested in diverse problems in Physics and her research area is theoretical Nuclear Physics.

This article discusses 'exoplanets' and the methods of exploring them. As the word speaks for itself, *exo* in Greek means *outside*. Hence exoplanets refer to extrasolar planets; every planet found outside our solar system is an exoplanet. An exoplanet can look

"Somewhere, something incredible is waiting to be known"
– Carl Sagan

like any planet of our solar system, but can also be a super-Earth, hot-Jupiter, hot-Neptune, cold-Jupiter, cold-Neptune, desert-planet, ocean-planet, mini-Neptune, ice-giant, gas-giant, super-puff, carbon-planet, lava-planet, Chthonian-planet and so on.

## 2. Voyage to Hunt Exoplanets

[1] See Sujan Sengupta, The Search for Another Earth–I and II, *Resonance*, Vol.7, No.7 and 10, 2016.

The journey that led to the confirmation of exoplanets[1] began almost a century ago when in 1917, the evidence of an exoplanet was found around a star called Van Maanen 2. However, it was confirmed only in 2016 and came out to be the earliest-known extraterrestrial planetary matter. In 1960, Dr Frank Drake performed an experiment called Project Ozma, which attempted to detect radio signals from extraterrestrial life and marked the beginning of Search for Extraterrestrial Intelligence (SETI). Following this was the development of the Drake equation, which predicts the number of active intelligent civilizations in our galaxy [4]. Finally, In 1992, the first two exoplanets—PSR B1257+12 c and PSR B1257+12 d—were confirmed around a pulsar PSR B1257+12 [2], which is a neutron star located 2300 light-years away. These two exoplanets are super-Earths and are also called the first pulsar planets.

Did you know that the 2019 Nobel Prize in Physics was awarded for the discovery of 51 Pegasi b?

The journey proceeded with the confirmation of a huge Jupiter-like planet 51 Pegasi b, orbiting a main-sequence star 51 Pegasi in 1995 [5]. Did you know that the 2019 Nobel Prize in Physics was awarded for the discovery of 51 Pegasi b? No doubt, back then, this was a colossal discovery. But it led to confusion among the scientific community due to the fact that Mercury, a small rocky planet of our solar system, takes 88 days to complete its orbit around the Sun, but 51 Pegasi b, despite being a giant hot-Jupiter, takes only 4.4 days to complete its orbit; hence, it is too close to its parent star. The first few discovered exoplanets were more like a puzzle; the massive exoplanets had shorter orbits with unexpectedly higher or lower eccentricities.

As we entered the 21st century, this field escalated rapidly with advancements in technology. Telescopes such as Hubble, Spitzer,

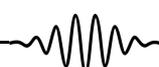

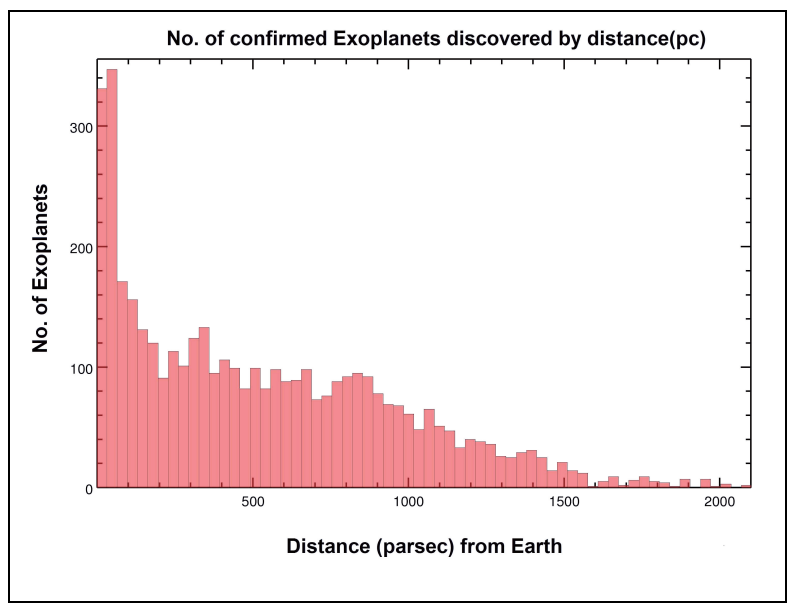

Kepler, K2, CoRoT, TESS, HARPS, WASP, etc., marked a new revolution in the field of extrasolar planets. In less than a decade, some of these telescopes identified the atmospheres of numerous exoplanets, ranging from chemical elements and compounds like sodium, water, carbon dioxide, and methane to prebiotic molecules that are probably the early building blocks of biosignatures and life. Humans might have thought in the past that our Earth is inimitable, but after years of speculation, here we have 4276 confirmed planets and 5481 NASA candidates[2] (as of 3rd September 2020 ) [6]. *Figure* 1 shows a histogram of the number of exoplanets discovered up to a distance of 2100 parsec (pc) from Earth. Clearly, the two highest peaks show that the maximum number of exoplanets have been confirmed at a distance of 100 pc.

**Figure 1.** A histogram depicting the number of confirmed exoplanets and their distance (pc) (as of 3rd September 2020).

[2] A NASA candidate is a likely planet discovered by different instruments, but whose existence and planetary information has not been verified yet.

## 3. Methodology of Detection

If exoplanets were always there, why did it take so long to discover them? This is actually pretty much self-explanatory. Finding an exoplanet is not just gazing at the night sky and discovering one. Usually, a planet is tough to detect because of the sun's glare.

An orbital zone around a star, also called the 'Goldilocks zone', where a planet can hold liquid water is called the 'habitable zone'.

But before moving any further, there is one important term to understand—the habitable zone (HZ). An orbital zone around a star, also called the 'Goldilocks zone', where a planet can hold liquid water, is called the 'habitable zone'. The possibility of occurrence of liquid water is very much high in the habitable zone, although it is not compulsory, and it can also occur outside the zone. However, this zone is not at a constant distance for all the stars. With the change in temperature of the sun, the zone changes its radius. The habitable zone is of two types:

**Optimistic Habitable Zone (OHZ):** The HZ where the recent Venus and early Mars were holding liquid water, as calculated and observed.

**Conservative Habitable Zone (CHZ):** The HZ where maximum greenhouse effect can occur.

As evident from *Figure* 2, we see the area of OHZ and CHZ of main-sequence stars changing and the value of effective stellar flux $S_{\text{eff}}$ is given on x-axis. Some of the exoplanets discovered in the HZ of their respective parent star are also shown. Here, $S$ is the solar flux, i.e., the radiant energy of the star reaching the exoplanet per unit area per unit time. $S_{\text{eff}} = S/S_o$ where $S_o$ is the normalized flux of Sun reaching the Earth [7].

Broadly, there are two methods to find exoplanets—-the direct method and the indirect method.

**Direct Method:** The direct method comprises direct imaging of a suspected planet orbiting a star with advanced telescopes. It uses adaptive optics, coronagraphs, etc., and a spectrum is also taken for spectroscopy. Though it is very promising, yet it is hard to detect exoplanets using direct imaging. Firstly, it becomes difficult to spot a planet in the star's light. Secondly, it works best only with infrared rays. That is why only a few exoplanets have been discovered using this method.

**Indirect Method:** Indirect methods look for the effects of exoplanets on the parent stars or the other nearby stars in the field [8]. This technique is the most used as it gives sensitive data and

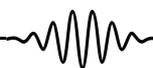

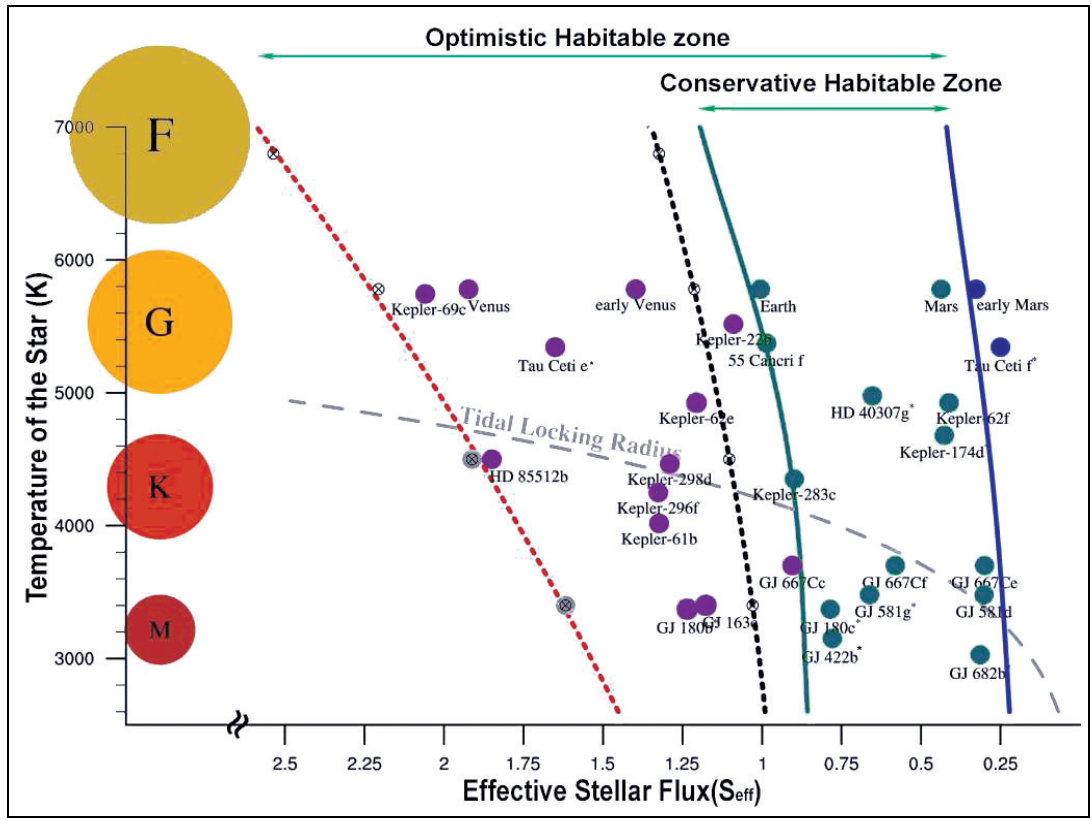

**Figure 2.** A schematic diagram representing the habitable zones of various main-sequence stars along with effective stellar flux of respective stars. The colored dots are the exoplanets found in the habitable zones of their parent star.

promising results. Some of the major types of indirect methods are described below:

### 3.1 *Transit or Eclipse Method*

A promising, flourishing, and widely used method is the transit photometry method. Whenever an exoplanet passes in front of its parent star during revolution, it transits and blocks a part of the light coming from the star. And that is where the science works! No matter how small the exoplanet is, there is always a dimming in the sun's light during the eclipse. Hence, the size and shape of the planet can be detected. The transit method is shown in *Figure 3*, where (a) is the light curve plotted between the star's observed brightness and time. The transit depth reaches a maximum when an exoplanet is completely in front of the star. This indicates the

> No matter how small the exoplanet is, there is always a dimming in sun's light during the eclipse.

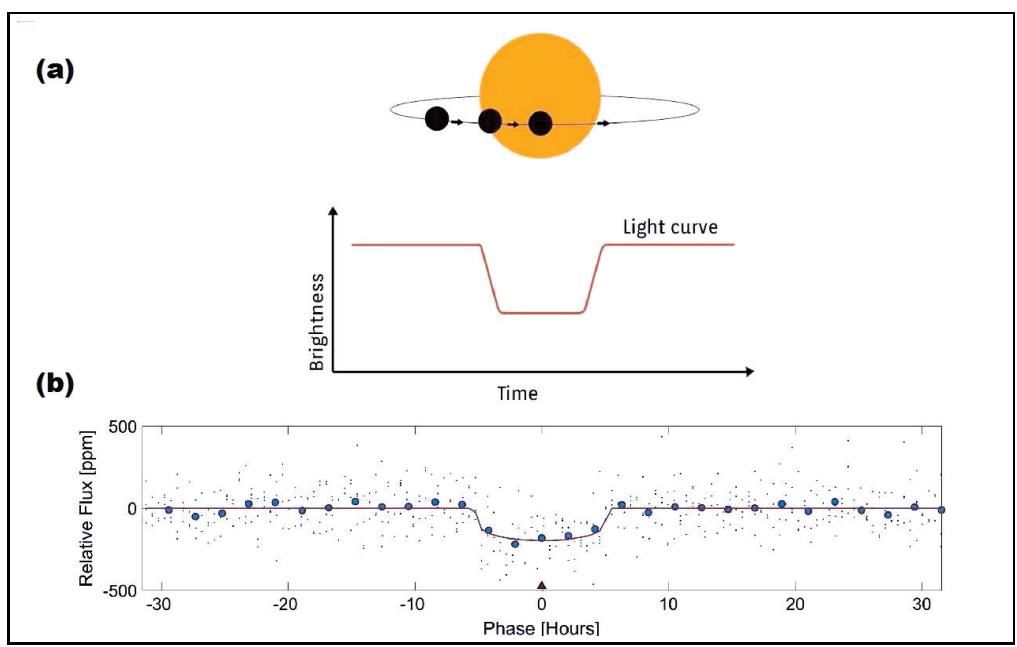

**Figure 3.** (a) Is the pictorial representation of transit method showing the transit depth because of the exoplanet crossing the star. (b) Is the face-folding light curve showing transit of Kepler-452b [9].

decrease in brightness during the eclipse. It is worth noticing that bigger planets have larger transit depths.

(b) in *Figure* 3 shows Kepler-452b exoplanet, 1402 light-years away, orbiting a G-type star Kepler-452. It was discovered by the Kepler space telescope in the constellation Cygnus by the transit method. Kepler-452b is often called Earth 2.0 because of it being a super-Earth, lying in the conservative habitable zone, having an orbit of 385 days and a potentially rocky nature [9].

### 3.2 *Radial Velocity (RV) Method*

This is one of the oldest methods that use high-resolution Doppler spectroscopy, as the basis of this method is Doppler shift[3] itself. Because of the mass of the exoplanet, it induces a wobbling in the star. If the exoplanet and the star are inclined along the sight of the observer, the star seems to move back and forth. This reflex motion caused by the exoplanet helps in its detection. When the star moves away from the observer, one sees a red-shifted spectrum, and when it moves towards the observer, one sees a blue-shifted

[3]A change in wavelength due to relativistic motion.

> When the star moves away from the observer, one sees a red-shifted spectrum, and when it moves towards the observer, one sees a blue-shifted spectrum.

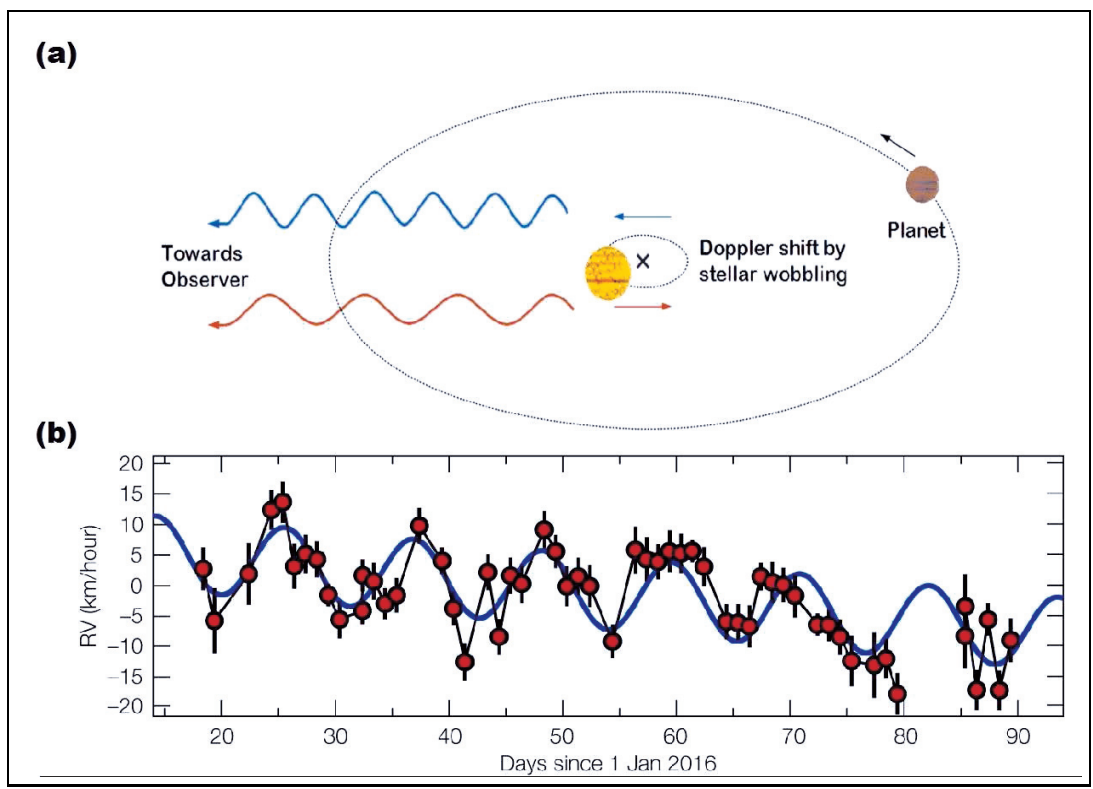

**Figure 4.** **(a)** Is a pictorial representation of Doppler shift occurring due to the wobbling of the sun because of the change in the direction of gravitational pull of the exoplanet. **(b)** Shows the radial velocity data curve of Proxima Centauri b recorded by HARPS.

spectrum. Therefore, this spectrum, combined with the calculation of the star's velocity, confirms the presence and mass of the exoplanet.

*Figure* 4 (a) is a pictorial representation of radial velocity showing the wobbling of a star because of a planet and its Doppler shift. From this figure, it can be concluded that there are two major factors affecting radial velocity variations. First, the exoplanet and star should be inclined in such a way that they are perpendicular to the observer's line-of-sight. Hence, RV is best observed when the angle of inclination is 90º and least when it is 0º. Second, the greater the mass of the exoplanet, the larger is the wobble of the star.

The bending of light due to gravity, such that the mass acts like a lens, produces gravitational lens effect.

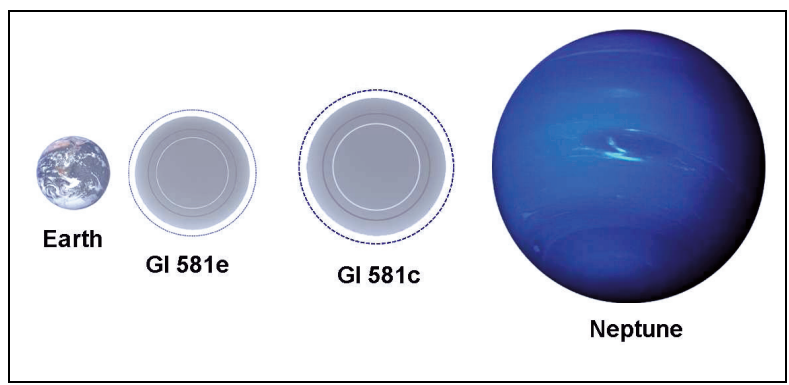

**Figure 5.** The size comparison of Gl 581e and Gl 581c with Earth and Neptune. These exoplanets were discovered by HARPS.

Using the radial velocity method, a planet called Proxima Centauri b [10] has been found. It is 4.2 light-years away, orbiting a red dwarf, in the constellation Centaurus. It is one of the nearest known exoplanets to Earth, having a mass of 1.7 times that of Earth and an orbit of 11.2 days around its star. *Figure* 4 (b) shows the radial velocity data curve of Proxima Centauri b taken by HARPS which shows the wobbling of the star and so the radial velocity.

### 3.3 *Gravitational Microlensing*

The bending of light due to gravity, such that the mass acts as a lens, produces the gravitational lens effect. Microlensing is used to detect low mass objects (exoplanets), and that's why it is observed on a human favourable time-scale—seconds to years. Usually, when there is a deviation in the standard picture of the sky, there is either a high magnification, or higher brightness observed because of microlensing, and hence an exoplanet is discovered.

The bending of light due to gravity, such that the mas acts like a lens, produces gravitational lens effect.

Two popular exoplanets Gl 581c and Gl 581e, found by HARPS using microlensing are orbiting a red-dwarf within Gliese 581 system. These exoplanets are two of the lightest known super-Earths [11]. *Figure* 5 shows the size comparison of both exoplanets with Earth and Neptune. However, besides Gl 581e and Gl 581c, there are two other planets (candidates) of this system, which are hoped to be in the habitable zone.

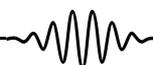

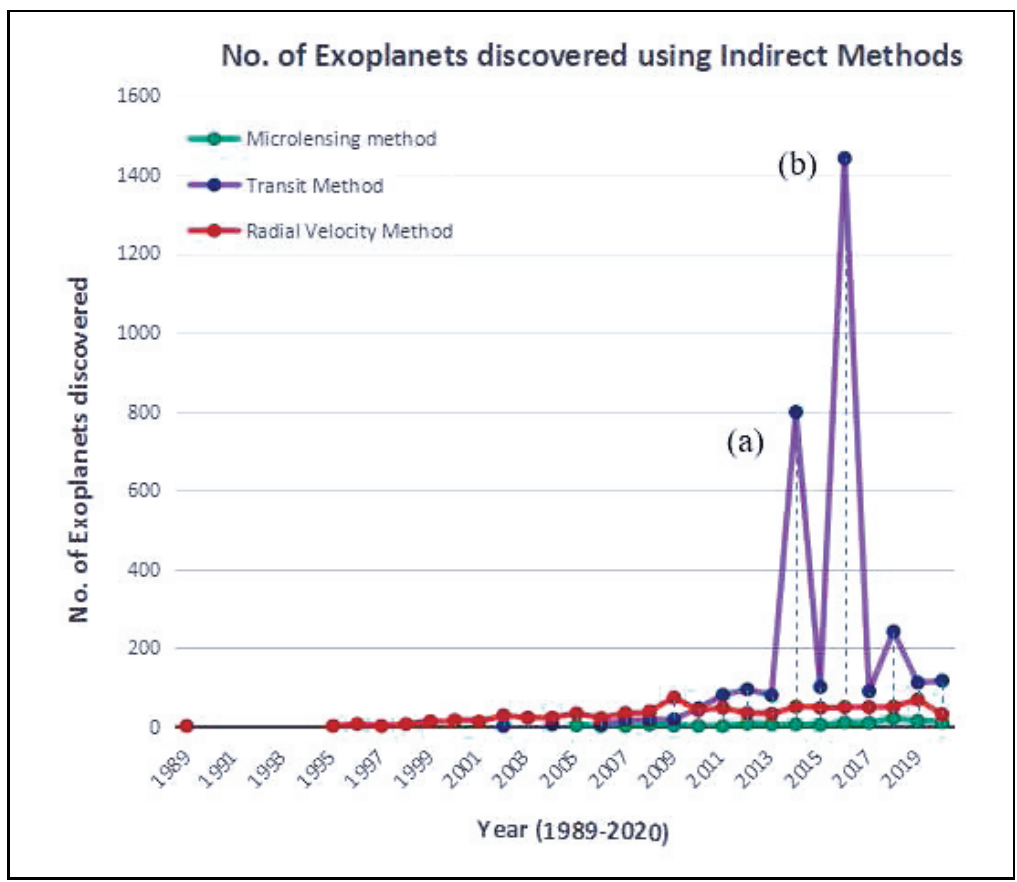

Along with these methods, there is the variable star method, pulsar timing method, transit timing method, polarimetry, interferometry, etc. All of these methods have their own pros and cons. Still, it is remarkable that most of the exoplanets have been discovered using the transit photometry method. Two significant peaks can be seen in *Figure* 6, which has never been observed by any other method yet. These two peaks show a rapid increase in the number of confirmed exoplanets—798 in 2014 (a) and 1442 in 2016 (b) [6].

**Figure 6.** A chart showing the number of exoplanets discovered using the indirect methods from 1989 to 2020. From this figure it can be concluded that the transit method has been the most successful method yet to find exoplanets.

## 4. Some Interesting Exoplanets

As the universe itself is so vast and mysterious, so are the exoplanets within it. With new technologies, scientists can apply various methods and spectroscopic techniques to study these planets. Here are some of the fun and astonishing exoplanets found so far.

- TOI 700 d: First Earth-sized planet with moderate temperature, orbiting its parent red dwarf in the habitable zone, 101.4 light-years away, discovered by TESS.

- Wasp-76b: A hot-Jupiter, orbiting an F-star, where iron rains.

- Kepler-444 system: The oldest planetary system of our galaxy having five terrestrial planets.

- Kepler-16b: Star Wars movie comes true! A planet of our galaxy where two suns set like the Tatooine world.

- 55 Cancri e: A planet with sparkling skies and boundless lava ocean.

- TRAPPIST-1: A system with 7 Earth-sized rocky planets; 3 of them in the habitable zone, hence potentially one of the best to observe.

- Gliese 504 b: A directly imaged planet, magenta in colour.

- K2-18b: A super-Earth, 124 light-years away, found to have an expressive amount of water vapour in its atmosphere.

> K2-18b is a super-Earth, 124 light-years away, found to have an expressive amount of water vapour in its atmosphere [12] [13].

## 5. Future Hopes from Exoplanets

Talking about aliens always seems fun! Though deep down, every human feels that there must be a single solid solution to the Fermi paradox, why haven't we found any alien civilization yet? [14] The Drake equation, that tells us how to estimate the number of extraterrestrial intelligent civilizations in our galaxy, calculated 1000 to 100,000,000 civilizations. In 2020, a paper has been published, assuming that there might be 36 communicating extraterrestrial intelligent (CETI) civilizations present in our galaxy based on astrobiological Copernican weak and strong limits[4] [15].

---

[4]The two astrobiological Copernican limits refer to the situation that intelligent life forms in less than 5 billion years or after about 5 billion years, similar to on Earth where a communicating civilization formed after 4.5 billion years.

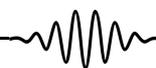

**GENERAL ARTICLE**

Although it is not physically possible to travel such a long distance (17000 light-years) in the upcoming years, we can assume what kind of life may exist on the exoplanets.

---

**Box 1. Some more interesting exoplanets that have surely blown everyone's mind.**

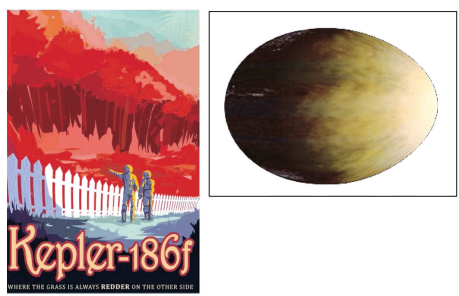

**Figure A.** **(a)** Is the grass always green? Well, apparently no! Kepler-186f is an Earth-sized rocky planet and the first to be discovered in the habitable zone of its parent star, a red dwarf Kepler-186 which is 582 light-years away from Earth. Because of the red dwarf's red wavelength photons, photosynthesis could have been affected thus making the grass there appear 'red' [16]. **(b)** Who said planets are only spherical? Here is WASP-12b, an 'egg-shaped' hot-Jupiter orbiting its parent star WASP-12, located 1411 light-years away. This exoplanet is so close to its host star that the star's gravity is tearing it apart and stretching it into an egg shape. Although it sounds fun, but this planet is being eaten by its star and will be completely consumed in 3.25 million years.

---

Let's take the example of Titan, one of the moons of Saturn where liquid methane oceans and methane rain is found. Although there has been no sign of life yet, there is a possibility of methanogenic life on it. In the same way, we can expect different types of lives in other worlds too. Complex biosignatures are hoped to be detected in the near future, though we don't know exactly what to look for. Any kind of technosignature would also be a great breakthrough, because among all the living species that originated on Earth, only human-life has been able to create technosignatures.

> Complex biosignatures are hoped to be detected in the near future, though we don't know exactly what to look for. Any kind of technosignature would also be a great breakthrough, because among all the living species that originated on Earth, only human-life has been able to create technosignatures.

Adding to the development of extraterrestrial planetary science, a very important and remarkable paper about exoplanets has been published in 2020 from India. It accounts for the study of transit timing variation of a hot-Jupiter, TrES-3b, located 1305 light-years away. 12 new transit light curves were reported based on observations using the Devasthal fast optical telescope at Devasthal observatory, ARIES, Nainital, India along with observations from the Indian Astronomical Observatory, Hanle, India, and Crimean Astrophysical Observatory, Nauchny, Crimea. It is worth mentioning that using these 12 transit light curves with the previous 71 transit light curves available in the literature, all the parameters[5] calculated were in agreement with previously calculated theoretical values [17].

Qatar-1 system is also an interesting extrasolar planetary system because of its strong transit signal and short orbital period. Transit timing variation analysis of 38 transit light curves has been done using HCT, Indian Astronomical Observatory, Hanle, India, which is also an outstanding contribution to the field of extraterrestrial planetary science by India [18].

Then there is PARAS (PRL advanced radial-velocity all-sky search) spectrograph, a very sensitive spectrograph mounted on a 1.2 m telescope in Mount Abu, Rajasthan, India. In 2018, the first exoplanet was detected by PARAS, a super-Neptune EPIC 211945201 b [19]. This exoplanet is 27 times the mass of the Earth, orbiting an F-type star and is 596 light-years away from Earth.

Besides these, with TESS, CHEOPS, JWST, PLATO, ARIEL, OUTLOOK, LU-VOIR, WFIRST, Gaia, and RST we might expect to find more earth-sized, rocky-planets in the habitable zones, with hopefully true biosignatures and technosignatures in the near future.

## 6. Summary

The ride for the search for exoplanets and extraterrestrial life goes through a tricky path. The field of extraterrestrial planetary science has grown exponentially and has resulted in knowledge about how exoplanets can look-alike and still be totally different at the same time. But, finding more and more intriguing facts about exoplanets could actually result in significant discoveries! Not only is there a chance to find intelligent civilizations like ours, but also scientists can learn about the formation of different essential chemicals and molecules (sometimes unexpectedly) even on what we consider the 'inhabitable' exoplanets. Hopes are high from upcoming exoplanet missions as we enter a new decade.

[5]Orbital parameters like ratio of planet to star radius, mid-transit time, orbital inclination, limb darkening coefficients, etc.

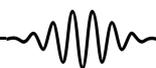

## Acknowledgments

We are highly thankful to the editor and anonymous referee for their generous comments that helped us to improve the article.

## Suggested Reading

[1] Eli Maor, To Infinity and Beyond: A Cultural History of the Infinite, Chapter 24, *The New Cosmology*, Birkhäuser Boston, USA, pp.198, 1987.

[2] A Wolszczan, D A Frail, A planetary system around the millisecond pulsar PSR 1257+12, *Nature*, Vol.355, No.653, pp.145–147, 1992.

[3] Frank D Drake, *Early SETI: Project Ozma, Arecibo Message*, https://www.seti.org/seti-institute/project/details/early-seti-project-ozma-arecibo-message, 1961.

[4] Frank D Drake, *The Drake Equation*, https://en.wikipedia.org/wiki/Drake_equation, 1961.

[5] Michael Mayor, Didier Queloz, A Jupiter-mass companion to solar-type star, *Nature*, Vol.378, No.6555, pp.355–359, 1995.

[6] NASA Exoplanet Archive, https://exoplanetarchive.ipac.caltech.edu/

[7] Ramses M Ramirez, A more comprehensive habitable zone for finding life on other planets, *Geosciences*, Vol.8, No.8, pp.280, 2018.

[8] Otto Struve, Proposal for a project of high-precision stellar radial velocity work, *The Observatory*, Vol.72, pp.199–200, 1952.

[9] Jenkins *et al.*, Discovery and validation of Kepler-452 b: A 1.6 $R_\oplus$ super-Earth exoplanet in the habitable zone of a G 2 star, *The Astronomical Journal*, Vol.150, No.2, pp.19, 2015.

[10] Guillem Anglada-Escudé *et al.*, A terrestrial planet candidate in a temperate orbit around Proxima Centauri, *Nature*, Vol.536, No.7617, pp.437–440, 2016.

[11] S Udry et al., The HARPS search for southern extra-solar planets-XI. super-Earths (5 and 8 M $\oplus$) in a 3-planet system, *Astronomy & Astrophysics*, Vol.469, No.3, pp.L43–L47, 2007.

[12] Björn Benneke et al., Water vapor and clouds on the habitable-zone sub-Neptune exoplanet K2-18 b, *The Astrophysical Journal Letters*, Vol.887, No.1, pp.L14, 2019.

[13] Angelos Tsiaras *et al.*, Water vapour in the atmosphere of the habitable-zone eight-Earth-mass planet K2-18 b, *Nature Astronomy*, Vol.3, No.12, pp.1086–1091, 2019.

[14] E Jones; NM Los Alamos National Lab (USA), *Where is Everybody: An Account of Fermi's Question*, United States 1985 03 01, 1985.

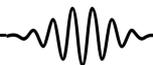

*Address for Correspondence*
Simran Kaur[1]
Varinderjit Kaur[2]
P G Department of Physics
G S S D G S Khalsa College,
Patiala, India.
Email:
[1]simrankaur33311@gmail.com
[2] drvarinderjit@gmail.com

[15] Tom Westby, Christopher J Conselice, The astrobiological Copernican weak and strong limits for intelligent life, *The Astrophysical Journal*, Vol.896, No.1, pp.58, 2020.
[16] Exoplanet Exploration: Planets Beyond Our Solar System, https://exoplanets.nasa.gov/alien-worlds/exoplanet-travel-bureau/
[17] Vineet Kumar Mannaday *et al*., Probing transit timing variation and its possible origin with 12 new transits of TrES-3b, *The Astronomical Journal*, Vol.160, No.1, pp.47, 2020.
[18] Parijat Thakur *et al*., Investigating extrasolar planetary system Qatar-1 through transit observations, *Bulletin de la Société Royale des Sciences de Liége*, Vol.87, Actes de Colloques, pp.132–136, Année 2018.
[19] Abhijit Chakraborty *et al*., Evidence of a sub-Saturn around EPIC 211945201, *The Astronomical Journal*, Vol.156, No.1, pp.3, 2018.

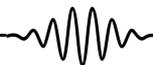